\documentclass[11pt,preprint]{aastex}
\usepackage{natbib}
\usepackage{color}
\usepackage{graphicx}
\usepackage{rotating}
\usepackage{hyperref}
\usepackage[all]{hypcap}
\usepackage{mathrsfs}
\usepackage{amsmath}
\usepackage{amssymb}
\usepackage{threeparttable}
\usepackage{enumerate}
\usepackage{bm}

\hypersetup{
    bookmarks=true,         
    unicode=false,          
    pdftoolbar=true,        
    pdfmenubar=true,        
    pdffitwindow=false,     
    pdfstartview={FitH},    
    pdftitle={My title},    
    pdfauthor={Author},     
    pdfsubject={Subject},   
    pdfcreator={Creator},   
    pdfproducer={Producer}, 
    pdfkeywords={keyword1} {key2} {key3}, 
    pdfnewwindow=true,      
    colorlinks=true,        
    linkcolor=blue,         
    citecolor=blue,         
    filecolor=magenta,      
    urlcolor=cyan           
}

\begin{document}
\title{The Magnetic Topology and Eruption Mechanism of a Multiple-ribbon Flare}
\author{Ye Qiu$^{1,2}$, Yang Guo$^{1,2}$, Mingde Ding$^{1,2}$, Ze Zhong$^{1,2}$}

\affil{$^1$ School of Astronomy and Space Science, Nanjing University, Nanjing 210023, China
\\ $^2$ Key Laboratory for Modern Astronomy and Astrophysics (Nanjing University), Ministry of Education, Nanjing 210023, China} \email{guoyang@nju.edu.cn; dmd@nju.edu.cn}

\begin{abstract}
Multiple-ribbon flares are usually complex in their magnetic topologies and eruption mechanisms. In this paper, we investigate an X2.1 flare (SOL2015-03-11T16:22) that occurred in active region 12297 near the center of the solar disk by both potential and nonlinear force-free field models extrapolated with the data observed by the Helioseismic and Magnetic Imager (HMI) on board \textit{Solar Dynamics Observatory} (\textit{SDO}). We calculate the three-dimensional squashing degree distribution. The results reveal that there are two flux ropes in this active region, covered by a large scale hyperbolic flux tube (HFT), which is the intersection of quasi-separatrix layers with a null point embedded in it. When the background magnetic field diminishes due to the separation of the northwest dipole and the flux cancellation, the central flux rope rises up forming the two brightest central ribbons. It then squeezes the upper lying HFT structure to generate further brightenings. This very energetic flare with a complex shape is accompanied by a coronal mass ejection (CME). We adopt the simplified line-tied force-balance equation of the current ring model and assign the observed value of the decay index to the equation to simulate the acceleration profile of the CME in the early stage. It is found that the path with an inclination of $45^\circ$ from radial best fits the profile of the actual acceleration.

\end{abstract}


\section{Introduction}

Flares are significant transient phenomena in the solar atmosphere. A large flare can release ~$10^{32}$ erg of free energy stored in the coronal non-potential magnetic field by magnetic reconnection. Flare ribbons are usually considered to be the product of accelerated electrons, which are produced by magnetic reconnection, propagate downward along the field lines, and heat the chromosphere or even deeper. Therefore, the morphology of flare ribbons is closely associated with the magnetic topology. A flare with two parallel ribbons is the type that is most commonly observed. Such a type can be well explained by the two-dimensional (2D) standard flare model \citep[also called the CSHKP model;][]{1964Carmichael,1968Sturrock,1974Hirayama,1976Kopp}. When a flux rope erupts due to the magneto-hydrodynamic (MHD) instability, it pushes the field lines enclosing the flux rope upward and stretches a vertical current sheet beneath the flux rope, where the field lines on both sides reconnect in succession. With the magnetic reconnection going on, two parallel ribbons appear as the footprints of the newly formed flare loops and depart perpendicularly from the magnetic polarity inversion line (PIL).

However, the morphology of multi-ribbon flares is much more complicated that cannot be interpreted by the 2D standard flare model, such as circular-ribbon flares \citep{2009Masson,2009Su,2012Wang,2016Janvier}. A typical circular-ribbon flare is characterized by a circular or semicircular ribbon with two bright kernels locating within and outside it, respectively \citep{2009Masson}. This unique morphology corresponds to the fan-spine topology of a magnetic null point \citep{1990Lau}. The null point usually appears in the corona with a closed circular polarity surrounding an opposite parasitic polarity \citep{2009Torok}. At the vicinity of the null point, the magnetic field can be described by the linear form of a position vector where the coefficient is the Jacobian matrix of the field \citep{1996Parnell,1996Priest}. The sum of the eigenvalues of the Jacobian matrix needs to be zero based on the divergence-free condition of magnetic field, which means that two of them have an identical sign or are conjugate imaginary. In the former case, we get a radial null point with field lines, defined by two identically signed eigenvectors, radiating into or away from the point. In the latter case, the field lines defined by two conjugate imaginary eigenvectors surround the null point in a spiral way, causing a spiral null point \citep{1996Priest}. In general, the plane defined by these two eigenvectors of both cases is a dome-shaped fan surface, while the rest singly signed eigenvector defines the direction of the spines passing through the null point. The spines are divided by the dome-shaped fan surface into the inner and outer ones. The fan separatrix intersects the lower atmosphere forming the circular ribbon, while the two bright kernels are the footprints of the inner and outer spines respectively. The outer kernel vanishes when the outer spine is open outwards, generating jets in the solar atmosphere \citep{2009Su,2010Pariat,2012Wang,2019Zhang}. The field lines passing through null points are preferential sites for gathering charged particles and forming current sheets where magnetic reconnection generally occurs. The null-point reconnection can be triggered by the shearing or rotational perturbations \citep{2007Potin,2007Potin_Galsgaard} or the eruptive flux rope beneath the fan surface \citep{b2013Sun,2015Yang,2017Li}.

In some cases, the outer ribbon stretches much longer than the footprint of the outer spine because of the presence of quasi-separatrix layers \citep[QSLs;][]{2009Masson,2015Yang}. QSLs are the regions where the linkage of field lines possesses great gradients, which is prone to induce magnetic reconnection \citep{1996Demoulin,1997Mandrini,2002Titov}. Separatrices are one special kind of QSLs where the linkage is
discontinuous. To quantify the position of QSLs, \cite{1996Demoulin} proposed the norm, $N$, of
the Jacobian matrix of the field line mapping. In applications, however, $N$ varies at the two footprints of the same field line. \cite{2002Titov} improved the concept of QSLs by introducing the squashing degree, $Q$, which is invariant along the field line. \cite{2012Pariat} have compared three numerical methods and found the optimal scheme for computing $Q$ within a three-dimensional (3D) box. In practice, QSLs are defined as 3D magnetic volumes of finite thickness where $Q$ is much greater than 2. The field lines within QSLs reconnect with neighboring lines, a process known as the slipping reconnection \citep{2006Aulanier,2007Aulanier,2009Masson}. Continuous reconnection occurs in the QSLs wrapping a flux rope, resulting the accumulation of magnetic helicity and energy of the flux rope \citep{2013Guo,2016Yang}. In observations, there is a correspondence between the position of flare ribbons and the intersection of QSLs with the lower boundary of the atmosphere \citep{2009Masson,2011Chandra,2015Yang,2019Zhong}. Thus, the morphology of flare ribbons is related to the configuration of QSLs.

Besides circular-ribbon flares, there are some other events with more complicated morphologies. \cite{2016Liu} has reported a series of X-shaped flares occurring in active region 11967, whose morphology is essentially consistent with the intersection of a hyperbolic flux tube (HFT) with the photosphere. A number of three-ribbon flares have been observed and studied \citep{2014Mandrini,2014Wang,2017Bamba,2017Sharykin,2019Zhong}. However, each of them has individual eruption characteristics due to its distinctive magnetic configuration. \cite{2014Wang} discussed two extended flares with three parallel ribbons and suggested a ``fish-bone-like'' topology for such events.    \cite{2014Mandrini} showed three irregular ribbons and interpreted them as the combined footprints of three asymmetric null-point structures. In another event studied by \cite{2017Bamba}, there were also three irregular elongated ribbons that were thought to be produced by the reverse magnetic shear. The shear of the flaring region was opposite to the majority of the active region. \cite{2017Sharykin} mentioned a third thin ribbon between two thick ones, the former of which was related to a small-scale magnetic arcade structure in the photosphere. \cite{2019Zhong} observed a transformation from a circular ribbon to three relatively straight parallel ribbons, which was caused by the formation of a bifurcated flux rope. In a word, considering the complexity of the magnetic field, it is difficult to establish a unified model to explain the formation of multi-ribbon flares of various kinds.

Flares are often accompanied by coronal mass ejections (CMEs), which can be driven by the flare reconnection or torus instability \citep[see review by][]{2018Green}. The torus instability \citep[TI;][]{2006Kliem} occurs because the strapping component of the background magnetic field over the flux rope decreases with height at a sufficiently rapid rate, which is described by the decay index $n$. In observations, eruptive flux ropes begin to rise up rapidly at the TI critical height \citep{2012Liu,2012Savcheva,2019Guo} while confined ones do not because the decay index is below the threshold for TI \citep{2010Guo,2011Cheng}. Moreover, there is a positive correlation between the velocity of CMEs and the decay index \citep{2012Xu,2017Deng}. In the simulations by \cite{2008Schrijver}, the TI model can indeed match the early phase of the eruption of filaments if input appropriate initial parameters. However, there are still few comparisons between the kinematics of the flux rope predicted by the TI model and the early kinematics of an actually observed CME. Therefore, we also tend to compute the acceleration profile of a CME dominated by TI and compare it with observations. By searching for the optimal parameters that can yield results best matching observations, we can furthermore estimate the direction along which the magnetic flux rope propagates. %

In this paper, we employ both the potential and nonlinear force-free field (NLFFF) models to infer the 3D coronal magnetic field of the X-class flare on 2015 March 11 and perform a detailed analysis of the eruption process. Besides, we compute the acceleration profile of the CME accompanied with this flare in the early stage using the TI model with the decay index actually observed. The paper is organised as follows. In Section 2, we describe the data for analysis and the extrapolation methods. In Section 3, we exhibit the evolution of some key parameters of active region 12297 on 2015 March 11, the extrapolated 3D coronal magnetic field, and the acceleration profile of the CME. Section 4 presents a discussion and a summary of the results.

\section{Observations and Methods} \label{sec:data}
\begin{figure}
\begin{center}
\includegraphics[width=1.0\textwidth]{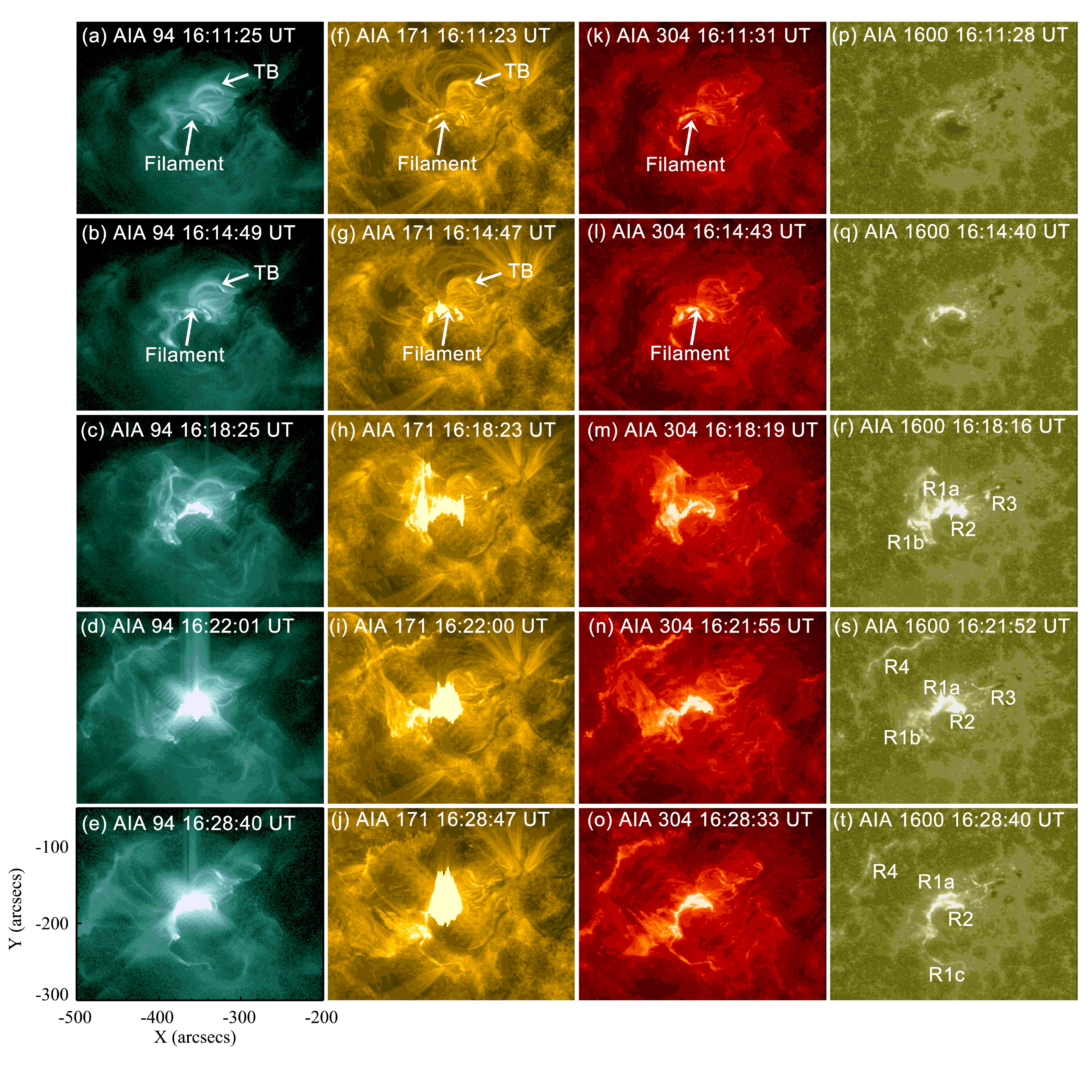}
\caption{ Multi-wavelength images of the multi-ribbon flare observed by \textit{SDO}/AIA. From left to right, the four columns present images at 94~\AA, 171~\AA, 304~\AA, and 1600~\AA, respectively. From top to bottom, each row shows images at nearly the same time. TB marks the tiny brightening before the eruption. R1a and R2 are the two brightest ribbons in the center. R1b is the J-shaped ribbon extending from R1a and followed by a circular ribbon R1c. R3 is an irregular ribbon in the northwest. R4 is the remote ribbon in the northeast.
} \label{fig:euv}
\end{center}
\end{figure}

Active region 12297 first appeared on the east limb of the solar disk on 2015 March 7 and rotated to the west limb on 2015 March 20, during which 105 flares occurred in succession in this region. However, among them only one reached X-class according to the record of the \textit{Geostationary Operational Environmental Satellite} (GOES). The X2.1 flare started at 16:11:00 UT, peaked at 16:22:00 UT, and ended at 16:29:00 UT on 2015 March 11. It was located at S$17^\circ$ E22$^\circ$ near the center of the solar disk. Initially, there was a tiny brightening to the northwest of a filament, as indicated by an arrow in the $94~$\AA ~and $171~$\AA~images (TB in Figure~\ref{fig:euv}(a),~\ref{fig:euv}(b),~\ref{fig:euv}(f) and~\ref{fig:euv}(g)). Then, the filament started to rise up at 16:14:49 UT (Figure~\ref{fig:euv}(b)), producing the two brightest ribbons R1a and R2 on either side of the filament (Figure~\ref{fig:euv}(r)). Ribbon R1a was circular in shape and extended to a small J-shaped ribbon eastward (R1b in Figure~\ref{fig:euv}(r)), which was simultaneously accompanied by a small-scale irregular brightening R3 in the northwest (Figure~\ref{fig:euv}(r)). There was an obvious separating motion between R1a and R2. After the filament reached a certain height, an elongated ribbon R4 appeared to the northeast of R1 for just a few minutes (Figure~\ref{fig:euv}(s)). Finally, when ribbon R4 weakened and the J-shaped brightening vanished, ribbon R1b continued extending counterclockwise to the south, forming a semicircular shape (R1c in Figure~\ref{fig:euv}(t)).

In this paper, we use data obtained by the different instruments on board the \textit{Solar Dynamics Observatory} \citep[\textit{SDO};][]{2012Pesnell}. The Atmospheric Imaging Assembly \citep[AIA;][]{2012Lemen} supplies full-disk images at seven Extreme Ultraviolet (EUV) and two Ultraviolet (UV) wavebands, with a pixel size of $0.6''$ and a cadence of 12 s and 24 s for EUV and UV, respectively. The higher-temperature wavebands, including 94 \AA ~(6.3 MK), 131 \AA ~(0.40 MK, 10 MK, 16 MK), 171 \AA ~(0.63 MK), 193 \AA ~(1.3 MK, 20 MK), 211 \AA ~(2.0 MK), and 335 \AA~(2.5 MK), typically show the features in the corona like flare loops, while the other lower-temperature wavebands, 304 \AA~(0.050 MK), 1600 \AA ~(0.10 MK), and 1700 \AA ~(continuum), are sensitive to the heating in the chromosphere and upper photosphere like flare ribbons.

The Helioseismic and Magnetic Imager \citep[HMI;][]{2012Schou} provides the photospheric vector magnetic field, which is inverted from the four Stokes parameters of Fe I 6173 \AA . The magnetic field data have a pixel size of $0.5''$ and a cadence of 720 s. We eliminate the $180^\circ$ ambiguity of the transverse component of the vector magnetograms, whose series name is \textbf{hmi.B\underline{ }720s}, by the ``disambig'' file and correct the projection effect since the location of active region 12297 deviates slightly from the disk center. Then, we perform a coalignment of the preprocessed magnetograms at 6:00 UT and 11:00 UT with that at 16:12 UT. After the above procedure, we select a region of $440''\times368''$ with $220 \times 184$ uniform grids, covering the active region, as the boundary to extrapolate the 3D coronal magnetic field.

We reconstruct the potential magnetic field using the Message Passing Interface Adaptive Mesh Refinement Versatile Advection Code \citep[MPI-AMRVAC;][]{2003Keppens,2012Keppens,2014Porth,2018Xia} with the Green function method \citep{1977Chiu}. As for the NLFFF extrapolation, we adopt the magneto-frictional method that was implanted into MPI-AMRVAC by \cite{a2016Guo}. The magneto-frictional method tries to solve the magnetic induction equation by iteration,

\begin{equation}
\begin{aligned}
\frac{\partial \boldsymbol{B}}{\partial t}=\nabla \times(\boldsymbol{v} \times \boldsymbol{B})+\eta \nabla^2 \boldsymbol{B}
\label{induction}
\end{aligned}
\end{equation}

\noindent where the velocity is calculated through a simplified momentum equation as $\boldsymbol{v} =\frac{1}{\nu} \boldsymbol{J} \times \boldsymbol{B}$ ($\nu$ is the magnetofrictional coefficient) and $\eta$ is the magnetic diffusion coefficient (see \citealt{a2016Guo} for more details). The iteration will stop when the numerical dissipation is small enough and the initial potential field is converged to an optimal NLFFF matching the vector magnetogram. Its applications to the \textit{SDO}/HMI data have successfully reproduced the non-potential magnetic structures such as sheared field lines, flux ropes, and magnetic null points when the initial conditions were set properly \citep{a2016Guo,b2016Guo}.

To investigate the magnetic topology of active region 12297, we compute the 3D distribution of the squashing degree, $Q$, which displays QSLs with finite thickness. The squashing degree is calculated by the diagonalization of the Jacobian Matrix, which describes the differential of magnetic field line mapping \citep{2002Titov}. \cite{2012Pariat} have compared the accuracy of three methods for computing the squashing degree and we adopt the third one of them (see Equations (12)--(22) in their paper).

There is another kind of photospheric vector magnetograms provided by HMI, which is named Space-Weather HMI Active Region Patches (SHARPs). The \textbf{hmi.sharp\underline{ }cea\underline{ }720s} data product has already disambiguated the transverse component of the magnetic field and been remapped to the center of the solar disk via the heliographic Cylindrical Equal Area (CEA) projection \citep{2014Bobra}. Besides, the components of the magnetic field ($B_r$, $B_\theta$, $B_\phi$) in the Heliocentric spherical coordinate obtained from this data product are identical to the components ($B_z$, $-B_y$, $B_x$) in the Heliocentric Cartesian coordinate \citep{a2013Sun}. We use this product to calculate the magnetic flux, the vertical current, and the ratio of direct current (DC) to return current (RC) in both positive and negative polarities.


\section{Results} \label{sec:ob_re}

\subsection{Magnetic Flux and Current Density} \label{sec:overview}

\begin{figure}
\begin{center}
\includegraphics[width=0.8\textwidth]{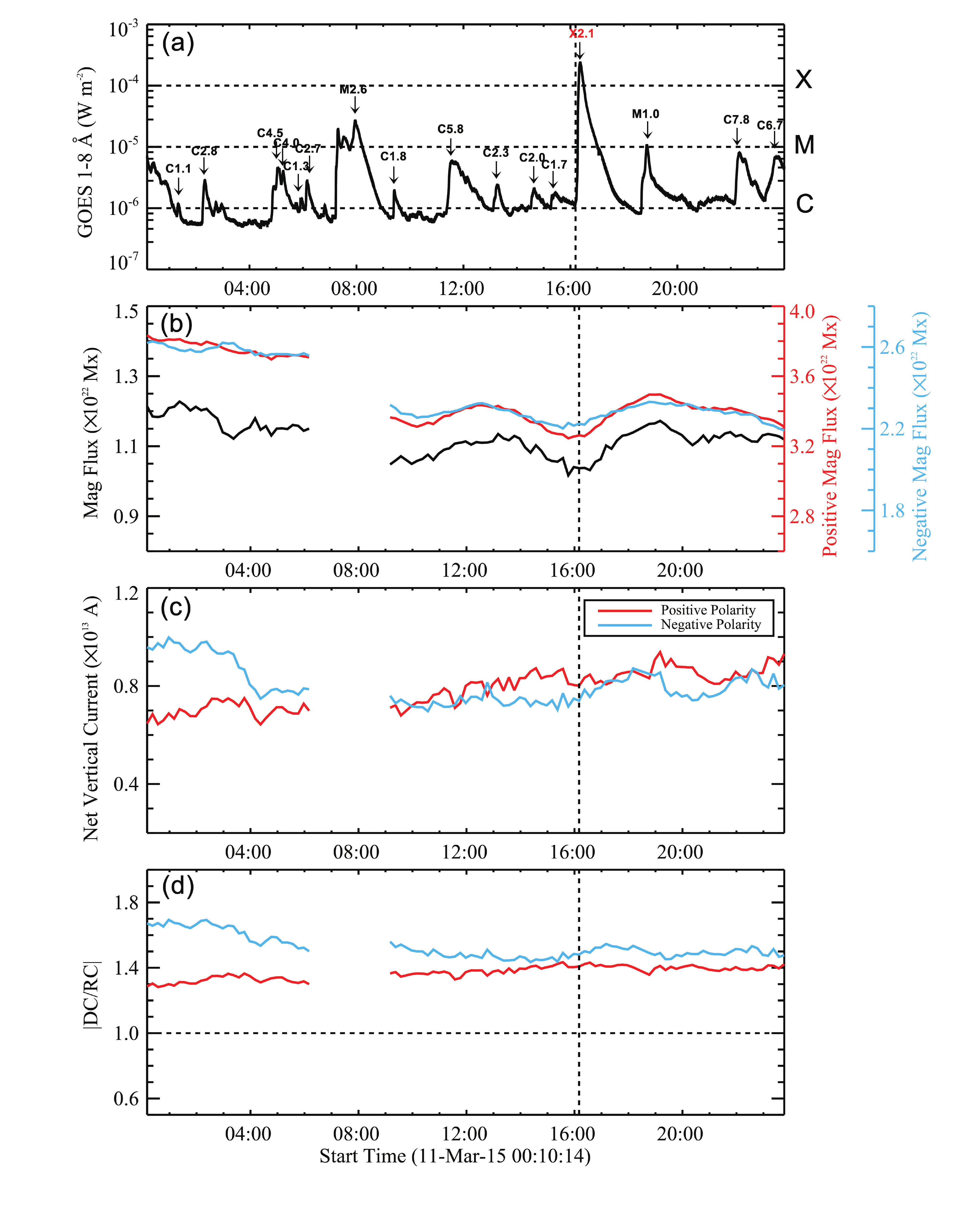}
\caption{Time evolutions of different parameters of the active region. (a) GOES Soft X-Ray flux. All the flares above C-class in active region 12297 during the whole day are indicated by the arrows. (b) Magnetic fluxes of the active region. Black, red, and cyan lines indicate the net flux, and the fluxes  in the positive and negative polarities, respectively. (c) Integrated photospheric vertical current. Red and cyan lines refer to that in the positive and negative polarities, respectively. (d) The ratio of direct current to return current. The line colors have the same meaning as in (c). All the physical quantities are presented from 00:10 UT to 23:46 UT on 2015 March 11. The vertical dashed line marks the start time of the X2.1 flare. There is a gap in SHARPs data from 6:10 UT to 9:10 UT as shown in (b)--(d). } \label{fig:all}
\end{center}
\end{figure}

Using the photospheric vector magnetogram obtained by SHARPs, we calculate the magnetic flux and vertical current of this active region. The time evolution of the net magnetic flux and the signed magnetic flux for each polarity are plotted in different colors (Figure~\ref{fig:all}(b)). The net magnetic flux in the whole region varies in the range of $(1.1-1.3)~\times~10^{22}$ Mx, while the signed magnetic flux fluctuates in the range of  $(3.3-3.9)~\times~10^{22}$ Mx and $(2.2-2.6)~\times~10^{22}$ Mx for the positive polarity and negative polarity, respectively. In general, the magnetic flux shows a decreasing trend with time, which means that active region 12297 has been decaying since 2015 March 11. However, the magnetic flux goes up for about 3 hours just after the start of the X2.1 flare, indicating that there is some new flux emerging into this active region during this period.

The current density is calculated by the Ampere’s law, $\mathbf{J} = \dfrac{1}{\mu_0}(\nabla \times \mathbf{B})$. Because the photospheric vector magnetograms are restricted to one layer, we can only calculate the vertical component of the current density via $J_z = \dfrac{1}{\mu_0}(\dfrac{\partial B_y}{\partial x}-\dfrac{\partial B_x}{\partial y})$. In this paper, we adopt three-point Lagrangian interpolation to increase the number of spatial grids and thus the accuracy in calculating the derivatives. Thus, we obtain the vertical current density in this active region on 2015 March 11. The net vertical current is calculated as the integration of the current density over the active region, i.e., $I=\int J_z~\mathrm{d} S$. If integrating $J_z$ of positive sign and that of negative sign separately, we can get the values of DC and RC, respectively, and finally the ratio of DC to RC. The DC and RC are the dominant and non-dominant components of the current in each polarity. In this active region, the dominant sign of the current density is identical to the sign of $B_z$. Thus, for positive (negative) polarities, the DC is obtained by integrating the positive (negative) current density, while the RC is the integration of the current density that has the opposite sign of $B_z$.

It is seen that, from 9:00 UT to 23:46 UT on 2015 March 11, the net vertical current in both polarities fluctuates in the range of $(6.9-9.3) ~\times 10^{12}~$A, suggesting that the current in this active region is approximately balanced (Figure~\ref{fig:all}(c)). However, there is an imbalance from 00:00 UT to 9:00 UT, which may be caused by the projection effects that cannot be removed totally by disk transformation when the longitude is too large. The ratio of DC to RC in positive and negative polarities undulates within 1.3--1.4 and 1.5--1.7, respectively (Figure~\ref{fig:all}(d)). This implies that active region 12297, which is CME-productive, deviates obviously from electric-current neutralization. This is consistent with the result of \cite{2017Liu} although the ratio $|\rm DC/RC|$ in our case is smaller than that in their cases.

\begin{figure}
\begin{center}
\includegraphics[width=1.0\textwidth]{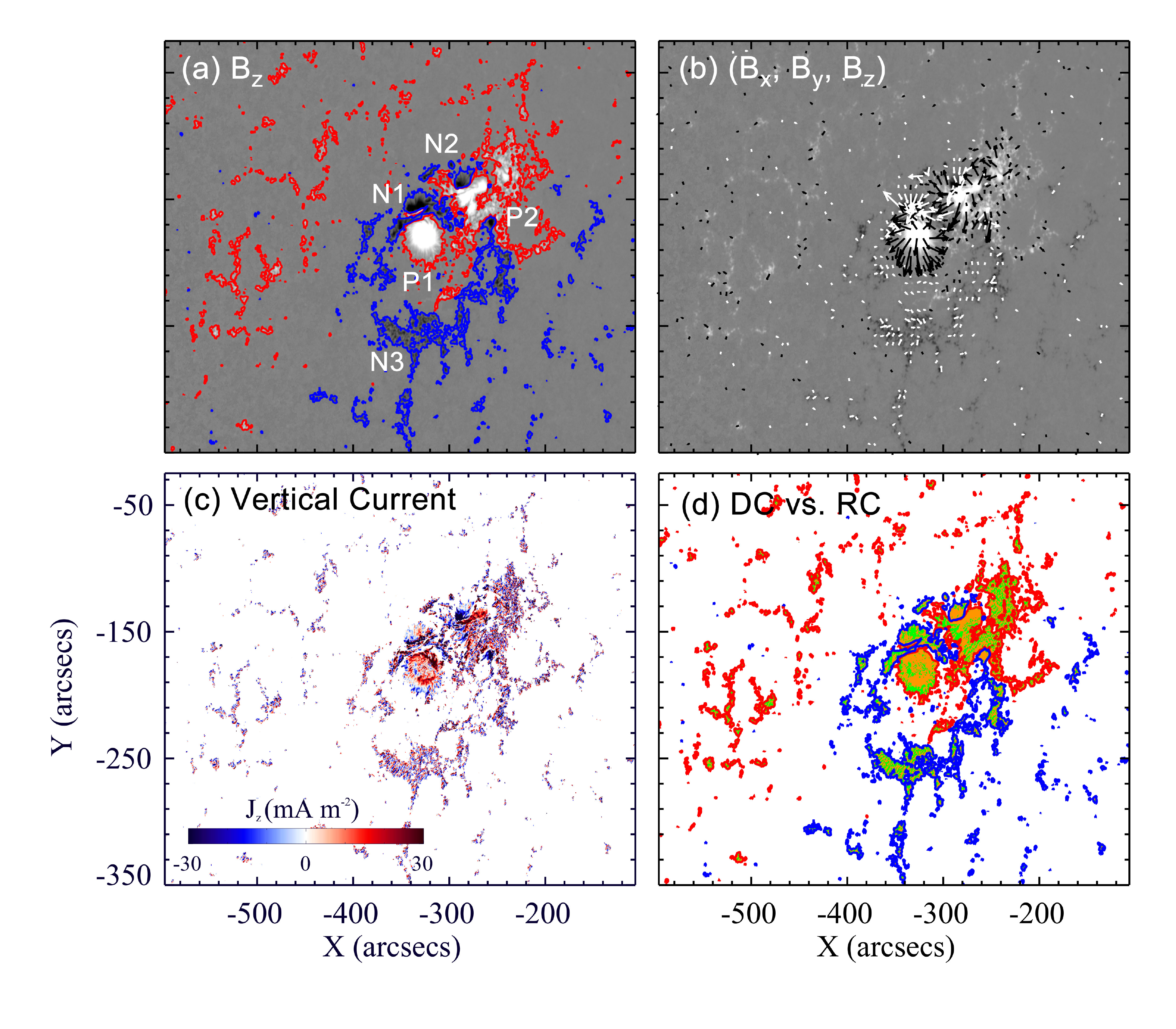}
\caption{Maps of photospheric magnetic field and vertical current of the active region at 16:10 UT. (a) Map of radial magnetic field component. The red and blue contours represent field strength of 150 G and $-$150 G, respectively. (b) Map of vector magnetic field. The horizontal component is shown as arrows. (c) Distribution of vertical electric current density. (d) Distribution of direct current (in orange) versus return current (in green). The colors of contours have the same meaning as in (a).
} \label{fig:current}
\end{center}
\end{figure}

Figure~\ref{fig:current}(a) and~\ref{fig:current}(b) shows the vector photospheric magnetic field of this active region at 16:10 UT on 2015 March 11. One can find that the morphology of magnetic field is very complex (Figure~\ref{fig:current}(a)). The two negative polarities, labelled as N1 and N3, constitute a circle surrounding the primary positive polarity, labelled as P1. To the northwest of N1 and P1, there is a dipole field that is composed of a small negative polarity N2 and a large diffused positive polarity P2. Figure~\ref{fig:current}(c) and~\ref{fig:current}(d) exhibits the vertical current density for this active region at the same time. Figure~\ref{fig:current}(c) reveals that the current has some coherent structures in the vicinity of PILs with in particular a double-J-shaped pattern between N2 and P2. In other areas, however, the electric current density is fragmented. Based on the signs of vertical current density and the signs of the corresponding magnetic polarities, we can draw a sign map exhibiting the distribution of the DC and RC (Figure~\ref{fig:current}(d)). The orange area refers to the DC density, which means a positive (negative) current density in a positive (negative) magnetic polarity. Likewise, the green area stands for the RC density, which implies a negative (positive) current density in a positive (negative) magnetic polarity. The result reveals that the DC is mainly distributed near the PILs but the RC is located away from it. In particular, in the central positive polarity P1, it looks like that the DC is surrounded by the RC.


\subsection{Initiation Mechanism} \label{sec:Initiation}

\begin{figure}
\begin{center}
\includegraphics[width=1.2\textwidth]{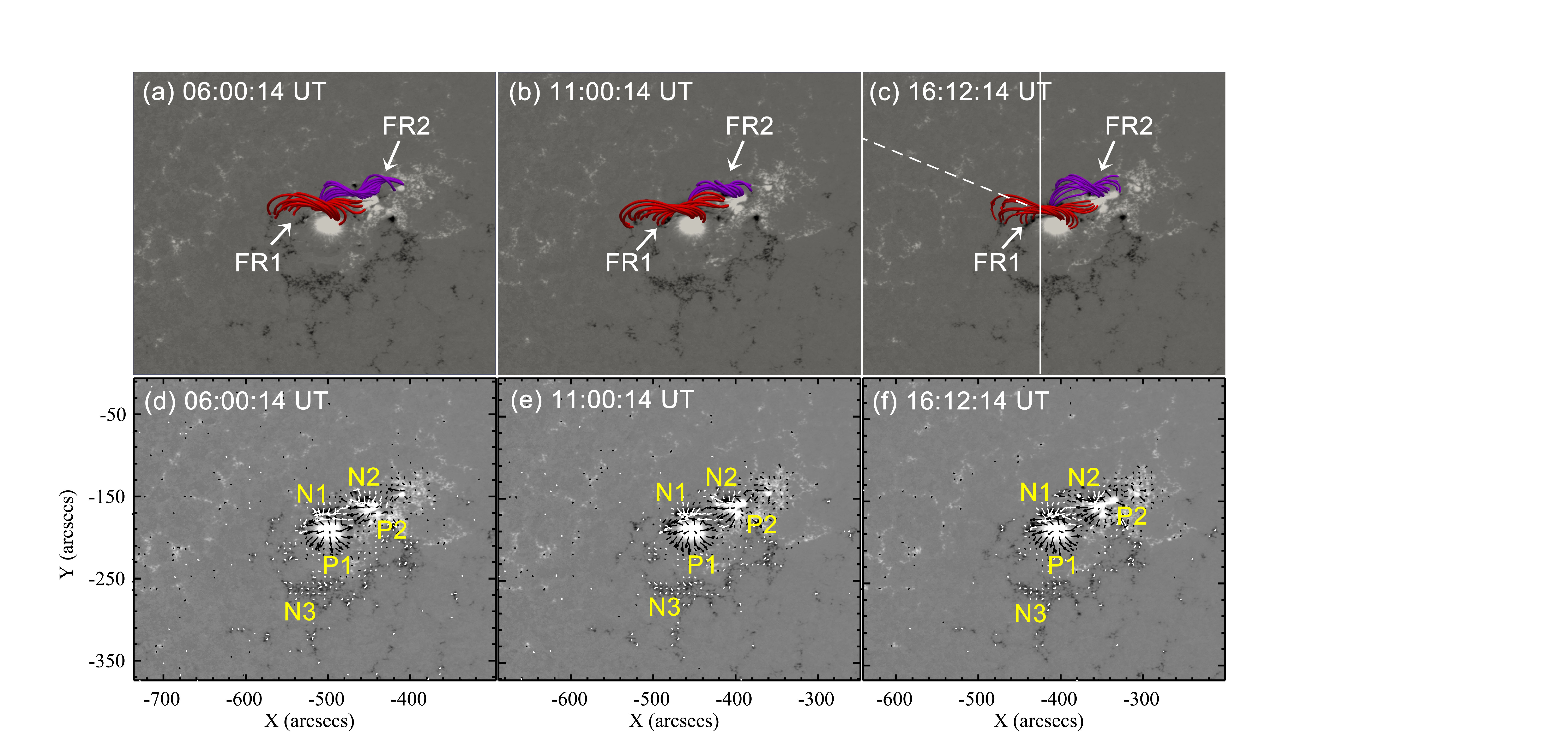}
\caption{(a)--(c) Top view of the configuration of two flux ropes at 06:00 UT, 11:00 UT, and 16:12 UT. FR1 is the central flux rope that erupts successfully to produce the X2.1 flare. The background represents the radial magnetic field. The white dashed line in (c) marks the position of the slice shown in Figure~\ref{fig:cme_fit}(a). The white solid line marks the slice for displaying the potential magnetic field in Figure~\ref{fig:back}. (d)--(f) Corresponding photospheric vector magnetic field. The radial field and transverse field are represented by the background and the arrows, respectively.
} \label{fig:pre}
\end{center}
\end{figure}

The flare under study is the only X-class flare occurring in active region 12297. To figure out its eruption mechanism, we need to learn the magnetic field associated with the flare. We coalign the vector magnetic field maps at 06:00 UT and 11:00 UT to that at 16:12 UT, which are then used as the boundary for extrapolating the 3D magnetic field. From the 3D magnetic field, one can see that two flux ropes, labelled as FR1 and FR2 (Figure~\ref{fig:pre}(a)), exist above the internal PILs of two dipoles (N1-P1 and N2-P2). The two dipoles move apart gradually with an obvious separating motion between N2 and P2. Consequently, the central flux rope (FR1) is more and more sheared to the PILs while FR2 gets untwisted gradually for about 10 hours after 06:00 UT (Figure~\ref{fig:pre}(a),~\ref{fig:pre}(b) and~\ref{fig:pre}(c)).

\begin{figure}
\begin{center}
\includegraphics[width=1.0\textwidth]{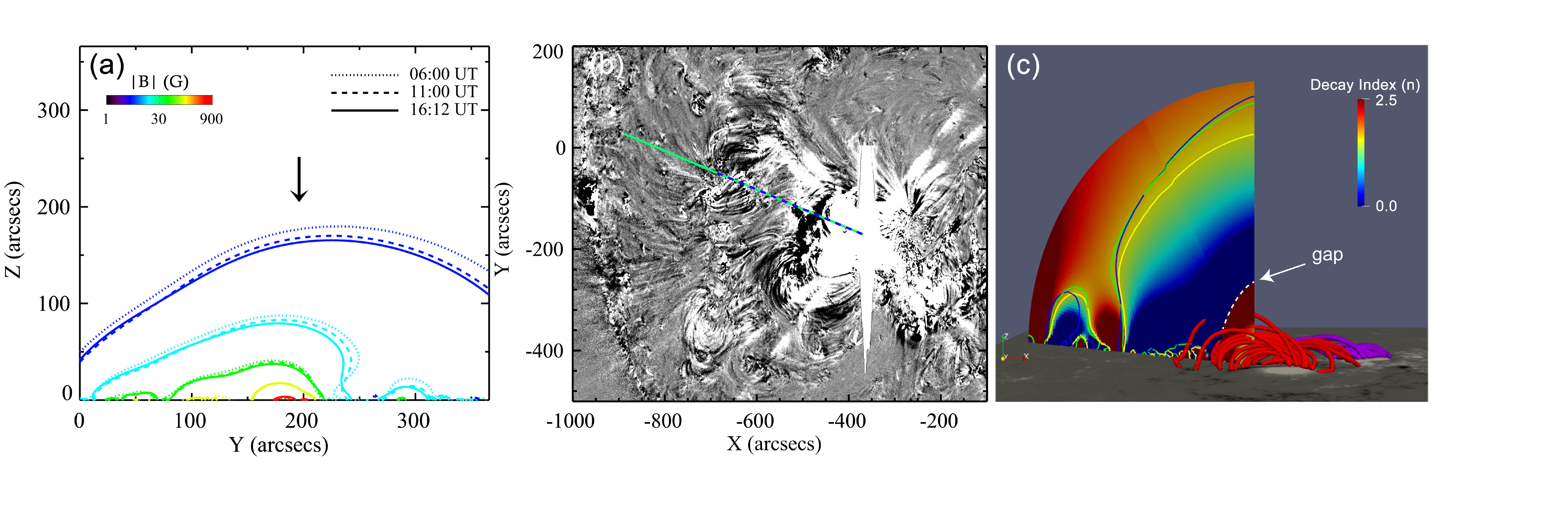}
\caption{(a) Contours of the potential field strength on the plane perpendicular to FR1 at three moments at $x=109$. The arrow points to the approximate position of FR1. (b) Base difference image of AIA 193~\AA~at 16:26 UT. The green solid line shows the slice used to trace the CME eruption in the plane of sky. The blue dashed line marks the ejection path projected to the solar surface. (c) The front view of the magnetic structure at 16:12 UT, superimposed with the distribution of the decay index at the same time (the colored sector). The dark blue, green, and yellow lines denote the contours at which the decay index is 1.5 at 06:00 UT, 11:00 UT, and 16:12 UT, respectively.
} \label{fig:back}
\end{center}
\end{figure}

We compute the potential field strength $|\textbf{B}_{\rm p}|$ in a vertical slice ($yz$-plane) at $x = 109$ grid (indicated by the white solid line in Figure~\ref{fig:pre}(c); corresponding to $-518''$ at 06:00 UT, $-470''$ at 11:00 UT, and $-422''$ at 16:12 UT, respectively). Such a slice is roughly perpendicular to FR1 and cuts though it in the middle. We then superimpose the contours of the field strengths at the three moments in one panel (Figure~\ref{fig:back}(a)). From the change of the iso-strength curves, one can find a decrease of the potential field with time. The weakening of the background field may be the result of the cancellation (Figure~\ref{fig:all}(b)) and separation (Figure~\ref{fig:pre}(d),~\ref{fig:pre}(e) and~\ref{fig:pre}(f)) between different magnetic polarities, or it is due to the rearrangement of the magnetic field caused by a number of smaller flares prior to the X-class flare. To understand the influence of the weakening of the potential magnetic field on the eruption of the X-class flare, we calculate the decay index along the CME propagation direction.

In the toroidal ring current model constructed by \cite{2006Kliem}, the hoop force (also named as the Lorentz self-force or expansion force) of a flux rope equilibrates with the stabilizing Lorentz force (restoring force) produced by the field lines passing over the flux rope. This equilibrium collapses owing to the torus instability provided that the restoring force attenuates faster than the expansion force with height when the flux rope expands. How fast the restoring Lorentz force by the external field decreases with height can be described by the decay index. In general, the decay index is given by $n = -\dfrac{\partial \ln B_{\rm pol}}{\partial \ln s}$, where $s$ is the ejection distance and $B_{\rm pol}$ is the poloidal component of the external magnetic field strength. Note that the key component hindering the eruption is the poloidal one perpendicular to both the flux rope axis and the propagation path. Here, we define the internal PIL between N1 and P1 as the axis direction of flux rope FR1 and denote its unit vector by $\textbf{e}_{\rm FR1}$. However, we cannot determine exactly the eruption direction of FR1 since there is no observation from another perspective, such as STEREO. Here, we try to infer the propagation direction of the CME on the bottom boundary of the extrapolated magnetic field. First, we trace the propagation path in the plane of sky from the AIA $193$~\AA~base difference maps (the green solid line in Figure~\ref{fig:back}(b)). Moreover, we project the path in the plane of sky to the solar surface (the blue dashed line in Figure~\ref{fig:back}(b)). Then, we convert the projected path to the Heliocentric Cartesian coordinate with the same way as that in the preprocessing of the magnetic field map at 16:12 UT. As a result, we can restrict the propagation direction in the $xy$-plane that is marked by the white dashed line in Figure~\ref{fig:pre}(c). The inclination of the ejection is still undetermined and thus treated as a free parameter in the range of $1^\circ$ to $90^\circ$ with $1^\circ$ intervals. For each degree, the unit vector of the propagation direction $\textbf{e}_{\rm pro}$ should be reassigned. Based on the definition of $B_{\rm pol}$, we require the poloidal unit vector $\textbf{e}_{\rm pol}$ to be vertical to both $\textbf{e}_{\rm FR1}$ and $\textbf{e}_{\rm pro}$, which means $\textbf{e}_{\rm pol}~\times~(\textbf{e}_{\rm FR1}~\times~\textbf{e}_{\rm pro})= 0$ and $B_{\rm pol} = \textbf{e}_{\rm pol}\cdot\textbf{B}_{\rm p}$ \citep{2019Guo}.

Figure~\ref{fig:back}(c) reveals that the decay index near FR1 increases sharply in a short distance along each assumed path, which is determined by jointly considering the propagation direction in the $xy$-plane and the inclination of the ejection (free parameter), as long as the inclination angle is large enough. One can see that the decay index changes from a fairly large value (the region in crimson) to an extremely small value (the region in mazarine) with an obvious gap in between (Figure~\ref{fig:back}(c)). This is because the poloidal component of the potential field approaches zero at that gap. After such a gap, the decay index increases monotonically with the distance away from FR1 along the assumed path. However, if the inclination of the ejection is small, i.e. the ejection path is nearly parallel to the bottom boundary, the decay index could show a fluctuation along the ejection path because of some small or dispersive magnetic polarities on the boundary.
We further superimpose the contours of the critical decay index, $n$ = 1.5, at 06:00 UT, 11:00 UT, and 16:12 UT (by mazarine, green, and yellow lines, respectively) in Figure~\ref{fig:back}(c). It is seen that, with time going on, the central flux rope FR1 gets closer and closer to the critical layer of TI, to say, it is more prone to eruption.

\subsection{Magnetic reconnection and flare ribbons} \label{sec:reconnection}
\begin{figure}
\begin{center}
\includegraphics[width=0.8\textwidth]{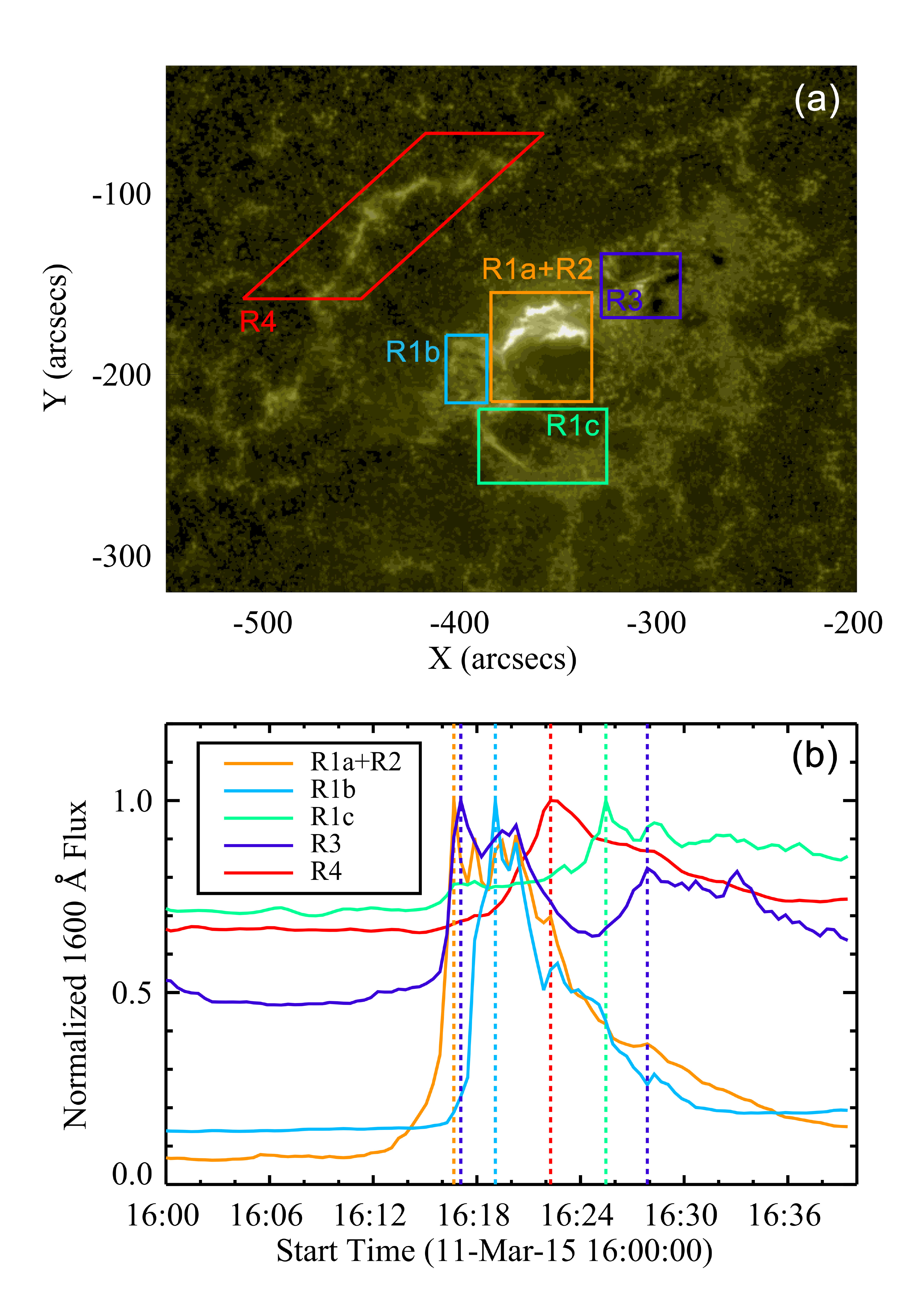}
\caption{(a) The 1600~\AA~image observed by \textit{SDO}/AIA at 16:27 UT, showing the brightening regions containing different ribbons. The boxes in different colors mark the regions we select to compute the 1600~\AA~flux. (b) Lightcurves at 1600~\AA~for various selected regions, normalized to their maximum value in each region. For each lightcurve, its peak time is indicated by a vertical dashed line.
} \label{fig:1600}
\end{center}
\end{figure}

We select five regions in the field of view of 1600~\AA~images in Figure~\ref{fig:euv} and integrate the UV flux in order to figure out the time sequence of discernible brightenings. We normalize the flux of each region to their maximum respectively, as shown in Figure~\ref{fig:1600}. The figure implies the following three stages of flare ribbon brightenings: (1) region R1a+R2 (the central ribbons) peaks at almost the same time as the first peak of R3 (the irregular ribbon in the northwest); (2) two minutes later, R1b (the J-shaped ribbon) reaches its maximum brightening, accompanied by a small peak of R3; and (3) the remote (R4) and southernmost (R1c) ribbons reach their emission peaks successively three minutes later, followed by the third peak of R3. Note that the ribbons brightening at the same time are supposed to be magnetically conjugated.

\begin{figure}
\begin{center}
\includegraphics[width=1.0\textwidth]{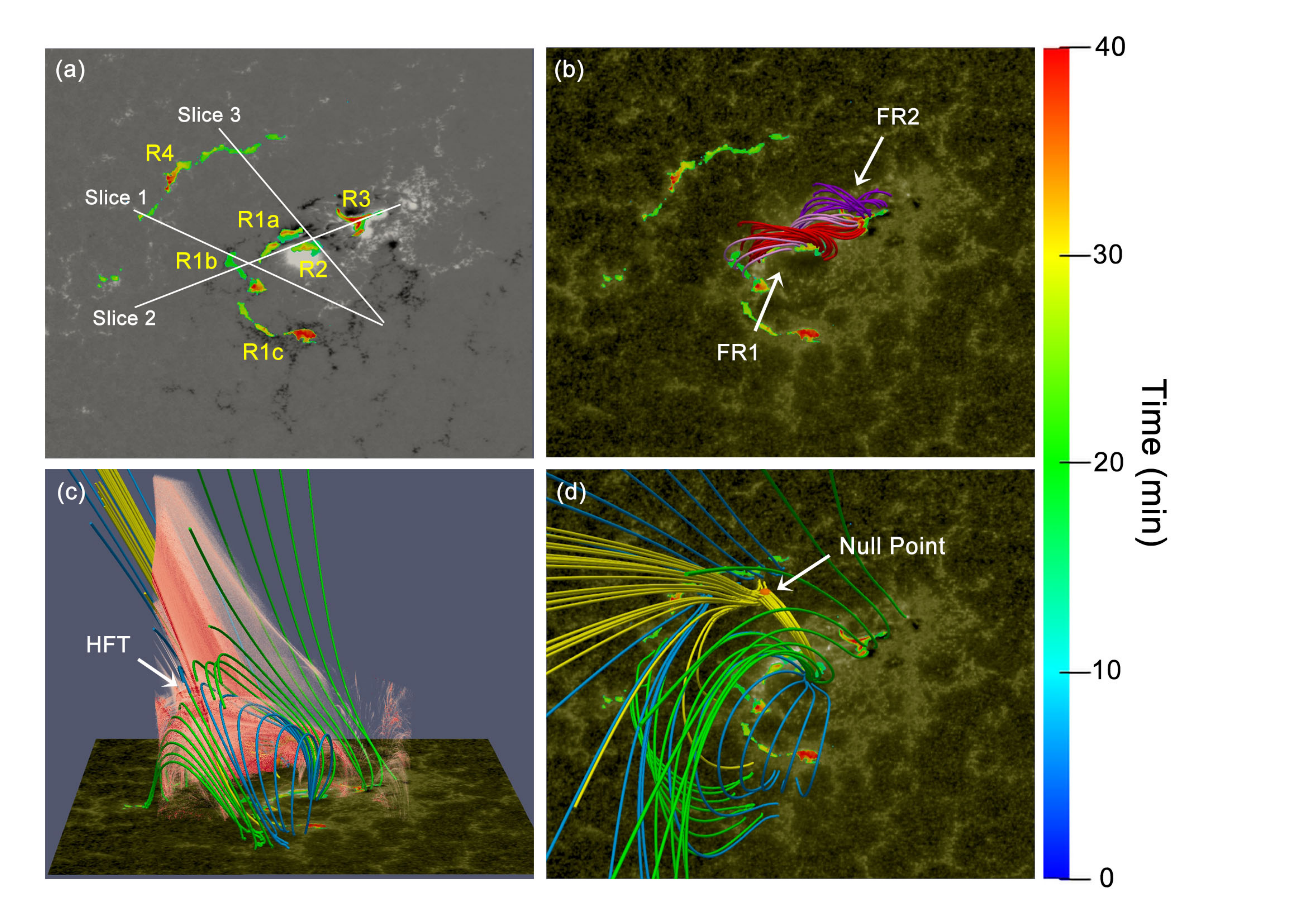}
\caption{(a) Magnetogram at 16:12 UT overlaid with the brightening sequence of flare ribbons. The white solid lines mark the projections of the three vertical slices used to show the $Q$-map in Figure~\ref{fig:qsl}. (b) Top view of the field lines constituting the flux ropes at 16:12 UT. The red and purple lines denote the flux ropes. The pink lines depict some slightly sheared field lines on the border of FR1. (c) Side view of the overlying magnetic structures. The three-dimensional halo structure is the large scale HFT formed by the intersection of a planar QSL and some arched QSLs (Figure~\ref{fig:qsl}). Some field lines close to the HFT are depicted as blue and green lines. The yellow lines pass through a null point. (d) Top view of the overlying field lines. The colors of the lines represent the same meaning as (c). The orange sphere indicated by the arrow shows the location of the null point.
} \label{fig:field_line}
\end{center}
\end{figure}

Figure~\ref{fig:field_line}(a) displays the evolution of the flare ribbons. One can see an obvious separating motion between R1a and R2. R3 slightly moves eastward while R4 drifts away from the center of the flare. Moreover, R1b and R1c seem to be rotating counterclockwise. We construct an NLFFF model using the vector magnetic field at 16:12 UT and exhibit some field lines overlaid on the 1600~\AA~image in Figure~\ref{fig:field_line}(b),~\ref{fig:field_line}(c) and~\ref{fig:field_line}(d). We find that FR1 is a bifurcated flux rope whose field lines connect one negative polarity but two positive polarities. Ribbons R1a and R2 are located on the negative and positive polarities N1 and P1, respectively, on both sides of FR1. The J-shaped brightening R1b corresponds to the footprint of the eastern part of the slightly sheared field lines enveloping FR1. Ribbon R3 is the combined footprint of the flux rope FR2 and the bifurcated flux rope FR1. Above FR1, there is a large scale HFT (in Figure~\ref{fig:field_line}(c)). We trace some field lines near the HFT. The footprints of these field lines are close to ribbons R1c and R4. We also find a null point (orange dot in Figure~\ref{fig:field_line}(d)) at a height of about 9.5 Mm defined by $|\mathbf{B}(x,y,z)| = 0$ and limited by the criterion as shown in the appendix of \cite{b2013Sun}. The outer spine stretches across the central flux rope FR1 and the inner spine is anchored close to the west part of R4.

\begin{figure}
\begin{center}
\includegraphics[width=0.8\textwidth]{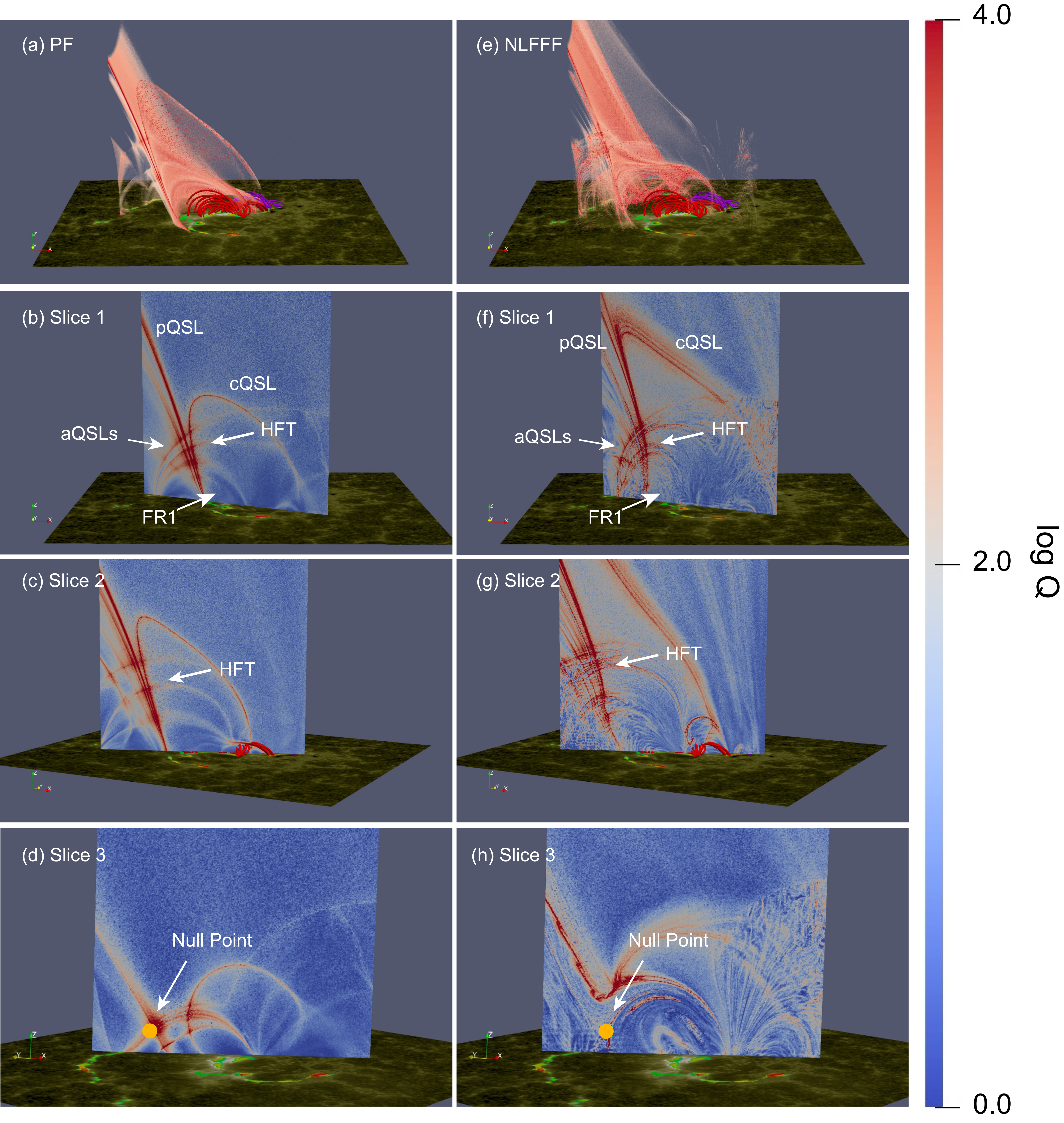}
\caption{(a)--(d) Distribution of the squashing degree, $Q$, computed from the potential magnetic field. (e)--(h) Similar to the left column but computed from the nonlinear force-free magnetic field. Note that pQSL is the planar QSL, cQSL is the cusp-shaped QSL, and aQSLs are some arched QSLs. FR1 points to the location of FR1. The orange dot in (d) and (h) marks the position of the null point. The background is the AIA 1600~\AA~image at 16:37 UT superimposed with the flare ribbons.
} \label{fig:qsl}
\end{center}
\end{figure}

The magnetic reconnection is prone to occur at QSLs, which are hard to visualize directly. Therefore we compute the squashing degree, $Q$, to depict QSLs in both the potential and NLFFF models. Figure~\ref{fig:qsl} presents the distribution of $Q$ in three-dimensional space and some vertical slices, which reveals that large squashing degrees are concentrated on three places forming three kinds of QSLs, one planar QSL, some arched QSLs and a cusp-shaped QSL. The planar QSL is inclined eastward on the east side of FR1, with these arched QSLs crossing it and the cusp-shaped QSL adjoining the western side of it. The arched QSLs and middle planar QSL constitute a large scale HFT, and their intersection rises from northwest to southeast. A part of FR1 lies below the HFT. When FR1 erupts, it may squeeze the upper lying HFT and induce magnetic reconnection at the HFT. From the slice along the spine, one can find that the null point is located at the vicinity of the HFT, even though the HFT is indistinct in the NLFFF model on slice 3 (Figure~\ref{fig:qsl}(d) and~\ref{fig:qsl}(h)).

According to the brightening sequence and magnetic topology of the active region, we surmise that there are three stages in the eruption of the FR1. Firstly, FR1 starts to rise up due to the weakening of the background magnetic field, causing the field lines enveloping FR1 to reconnect below, which forms the central two quasi-parallel ribbons (R1a and R2). Secondly, when FR1 rises, some slightly sheared field lines on the border of FR1 continuously reconnect with some field lines that connect polarities N1 and P2 in FR1 (pink lines in Figure~\ref{fig:field_line}(b)), as revealed by the 1600 \AA~emission  peak of R1b and the synchronous second emission peak of R3. Thirdly, FR1 further squeezes the large scale HFT and makes the overlying field lines reconnect sequentially in the vicinity of the HFT, resulting in the formation of the remote ribbon R4 and the circular brightening R1c. In addition, the circular brightening R1c is close to the southern circular part of the planar QSL footprints. As the magnetic reconnection at the HFT center occurs sequentially from bottom to top, R1c rotates counterclockwise from the east to the west.

\begin{figure}
\begin{center}
\includegraphics[width=0.8\textwidth]{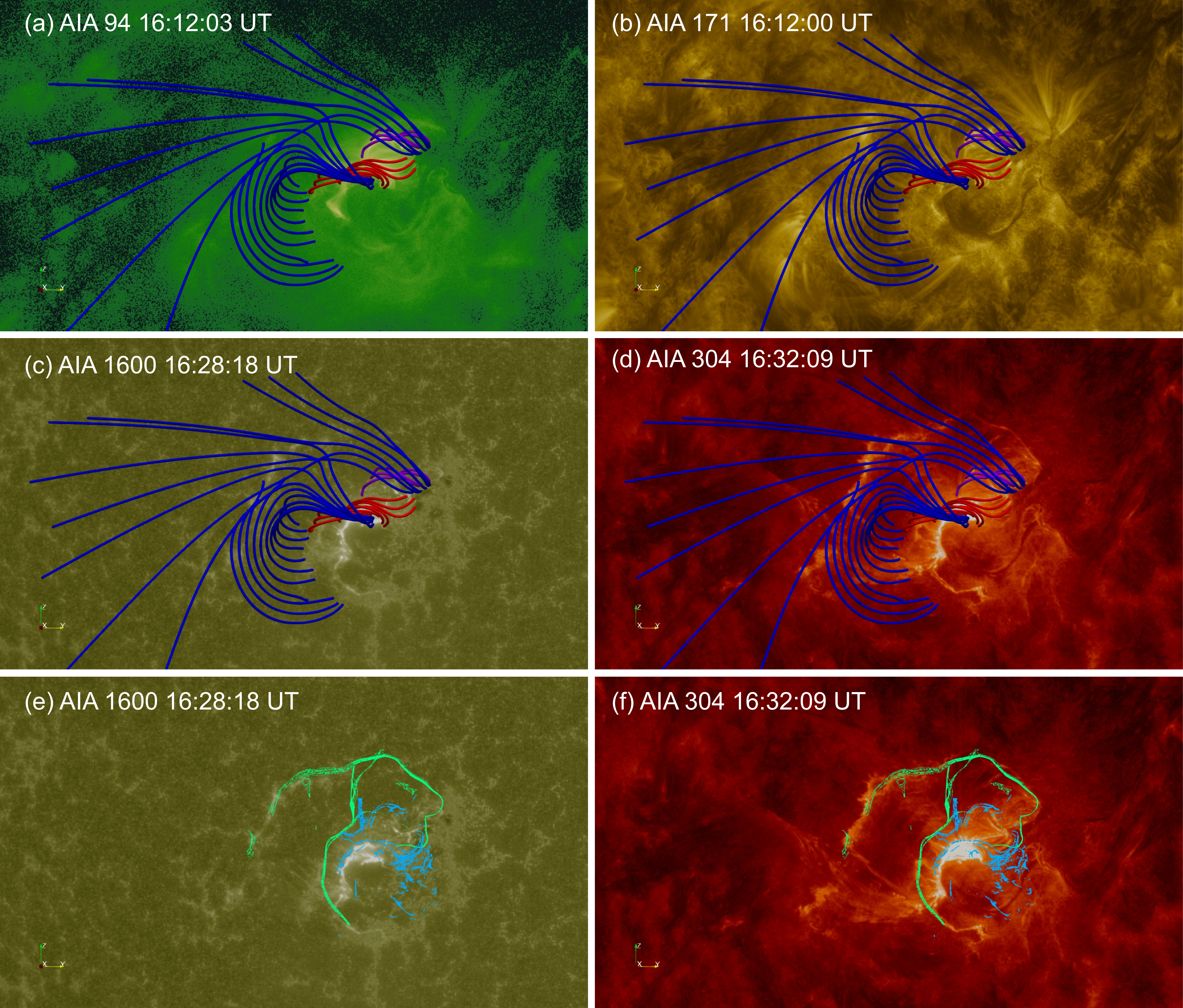}
\caption{(a)--(d) Comparison between AIA EUV/UV images and the extrapolated coronal NLFFF. The red and purple lines delineate the same flux ropes as that in Figure~\ref{fig:field_line}. The dark blue lines refer to magnetic field lines at higher altitudes. (e)--(f) Comparison between flare ribbons and QSLs on the $z=1.4 \rm~Mm$ plane. The green area represents QSLs from the potential field, while the blue area refers to QSLs from the NLFFF.
} \label{fig:pro}
\end{center}
\end{figure}

We project the NLFFF lines and QSLs to the perspective of AIA for a better comparison with the AIA observations (Figure~\ref{fig:pro}). One can see that the filament coincides with FR1 while the tiny brightening shown in Figure \ref{fig:euv}(a) coincides with FR2. The overlying field lines correspond to some higher coronal loops. The potential field QSLs are in accordance with the remote ribbons R4 and R1c as especially shown in the 304~\AA~image. Comparatively, the NLFFF QSLs are more associated with the brightenings near the flux ropes.


\subsection{The Acceleration Process of the CME} \label{sec:cme}
\begin{figure}
\begin{center}
\includegraphics[width=1.0\textwidth]{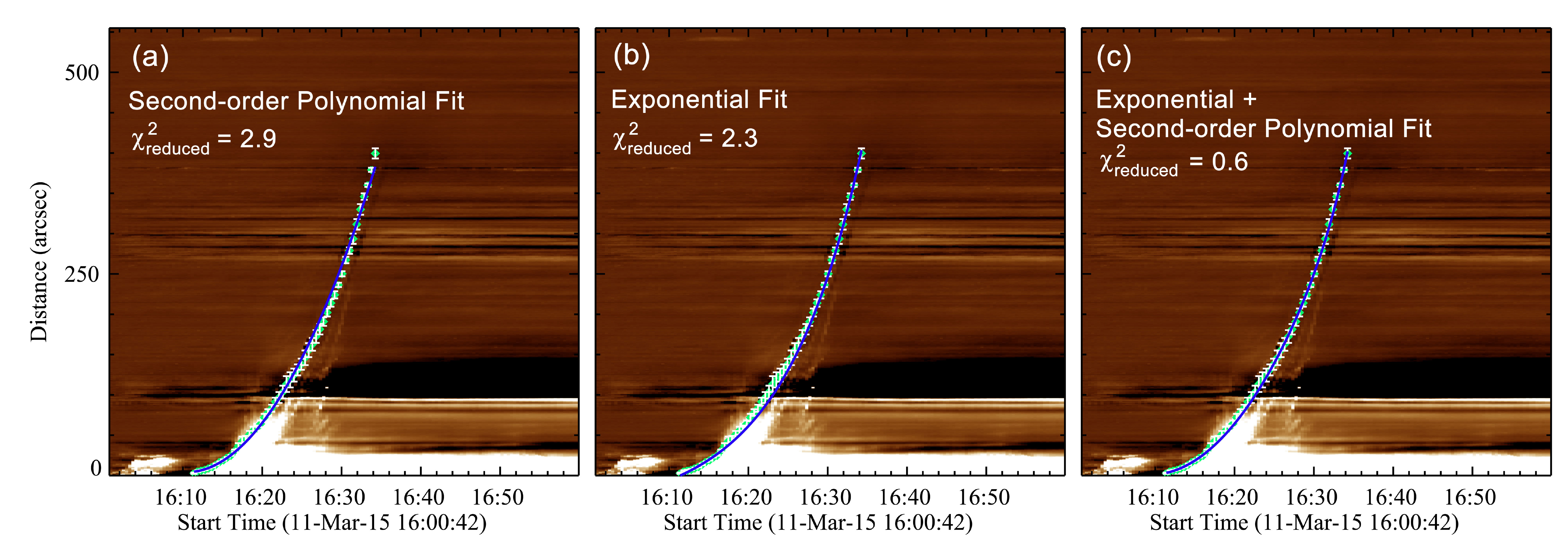}
\caption{(a)--(c) Fittings of the CME time-distance profile with three forms of functions. The green diamonds are measured points with vertical bars representing errors in measurement. The blue lines are the fitting curves.
} \label{fig:cme_fit}
\end{center}
\end{figure}

The X2.1 flare was accompanied by a CME that was observed by the Large-Angle Spectroscopic Coronagraph (LASCO) onboard the \textit{Solar and Heliospheric Observatory} (\textit{SOHO}) at 17:00:05 UT. To study the early acceleration of the CME, we make a slice (the green line in Figure~\ref{fig:back}(b)) along its propagation direction on the base-difference maps of AIA 193~\AA. The time-distance map along the slice shows that the CME, or the central flux rope, rises rapidly after the onset of the flare. To trace the kinematics of the CME, we select 50 time grids with intervals of 28 s and measure the position of the CME front at each time grid manually. The same procudure is repeated 5 times. We then make an average of the measurement as the distance that the flux rope has swept, while the standard deviation is regarded as the uncertainty in measurement. The position of the CME is indicated by the green diamonds with error bars overlaid on the time-distance maps (Figure~\ref{fig:cme_fit}). We use three different functions to fit the time-distance curve of the CME: a second-order polynomial function, an exponential function, and a combination of them. The fitting results and reduced chi-square, $\chi^2_{\rm reduced}$, are also displayed in Figure~\ref{fig:cme_fit}. The reduced chi-square is defined as $\chi^2_{\rm reduced} = \frac{\chi^2}{\rm DOF}$, where DOF is the degree of freedom and calculated by subtracting the number of the free parameters from the number of data points. The closer $\chi^2_{\rm reduced}$ is to unity, the better the fitting is. One can see that all the functions can fit the observation fairly well. By comparison, the fitting errors of the second-order polynomial function and the exponential function are relatively large, which means that the ejection of the flux rope deviates from a solely uniform acceleration or a singly explosive process.

\begin{figure}
\begin{center}
\includegraphics[width=0.9\textwidth]{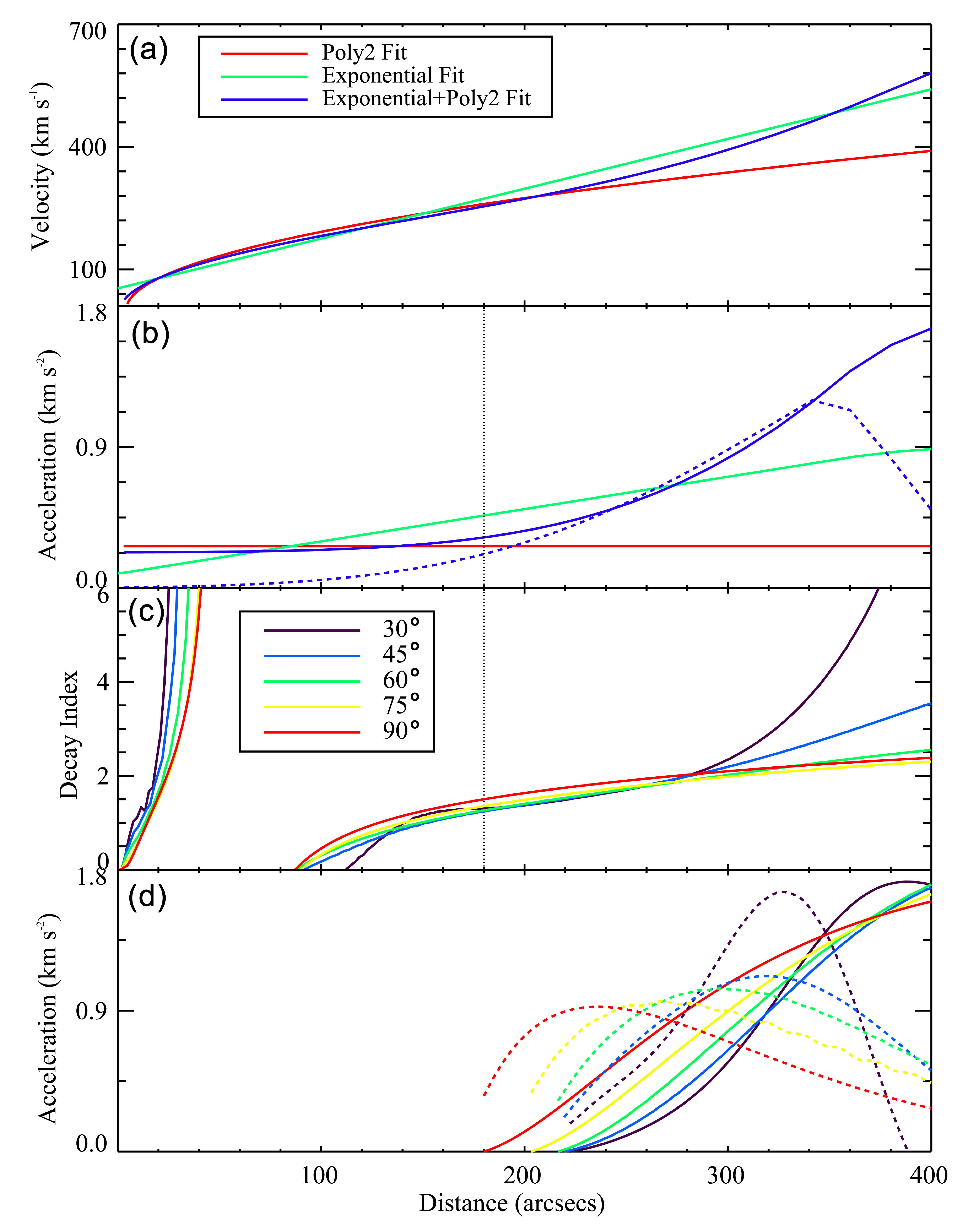}
\caption{(a) CME velocity as a function of the propagation distance derived by three fitting methods. (b) The acceleration of the CME derived from the velocity shown in (a). The dashed line displays the derivative of the acceleration with respect to the propagation distance in the third fitting case. (c) Variation of the decay index with the propagation distance when the CME is ejected at different directions. The vertical dotted line in (b) and (c) marks a propagation distance of 180$''$ where the decay index is close to 1.5 and the acceleration in the third case starts to increase rapidly. (d) The simulated acceleration and its derivative with respect to the propagation distance (dashed lines) in cases of different ejection directions.
} \label{fig:cme_curve}
\end{center}
\end{figure}

We further compute the velocity and acceleration from the three fitting cases and plot them against the propagation distance in the plane of sky in Figure~\ref{fig:cme_curve}. In the first fitting case, the velocity of the CME increases monotonously with an average acceleration of $0.27~\rm km~s^{-2}$. The acceleration in the second case rises from $0.10~\rm km~s^{-2}$ to $0.89~\rm km~s^{-2}$ with a linear dependence on the propagation distance, as expected. In the third case, the acceleration shows a quite different evolution: it remains at a relatively low value of $0.23~\rm km~s^{-2}$ or rises slowly at the beginning, but it then increases sharply to approximately $1.7~\rm km~s^{-2}$ after a propagation distance of about 126 Mm. As shown by the value of $\chi^2_{\rm reduced}$, this case is the closest fitting of the observed CME profile.

We extrapolate the potential magnetic field in a larger region of $864\arcsec \times 864\arcsec$ with the photospheric magnetogram at 16:12 UT. Then, we calculate the decay index towards five directions, whose inclination angles are $30^\circ$, $45^\circ$, $60^\circ$, $75^\circ$, and $90^\circ$ respectively, using the same method as that in Section~\ref{sec:Initiation}. By considering the projection effect, we can depict the curve of the decay index with the propagation distance in the plane of sky as viewed from the perspective of SDO. Note that the gap in Figure~\ref{fig:cme_curve}(c) is located at the same position of the gap in Figure~\ref{fig:back}(c), caused by the zero poloidal magnetic field but not QSLs. The location of the threshold $n$ = 1.5 is close to a propagation distance of about 180$''$ (126 Mm) in the plane of sky, where the acceleration of the CME increases rapidly, as revealed by Figure~\ref{fig:cme_curve}(b) and~\ref{fig:cme_curve}(c) (the third fitting case). It seems that the torus instability sets in and drives the eruption of the flux rope at this location. Based on this fact, we adopt the acceleration function under the assumption of line tying \citep[Equation (8) in][]{2006Kliem} to simulate the acceleration of the flux rope in the early stage,

\begin{equation}
\begin{aligned}
\frac{d^{2} \xi}{d \tau^{2}}=& \frac{1}{2\left(c_{0}+1 / 2\right)}+\frac{(2 n-3) c_{0}+1 / 2}{2(n-2)\left(c_{0}+1 / 2\right)}
\xi^{-1} - \frac{2 n-3}{2(n-2)} \xi^{1-n},~n \neq 2
\label{equation}
\end{aligned}
\end{equation}
where $\xi$ and $\tau$ are the normalized distance and time, $n$ is the decay index, and $c_{0}$ is the initial dimensionless inductance of the flux rope. The normalized distance and time are expressed as $\xi~=~R/R_{0}$ and $\tau~=~t/T$, where $R_{0}$ is the initial height of the flux rope, and $T = {\left(c_{0}+1 / 2\right)^{1 / 2} r_{0}}/{2V_{\mathrm{A} }}$ ($r_{0}$ is the minor radius of the flux rope and $V_{\mathrm{A}}$ is the $\rm Alfv\acute{e}n$ speed at the initial height). Because the flux rope is inherently unstable in the model of \cite{2006Kliem}, we set the initial height at which $n$ starts to exceed the TI threshold. Correspondingly, we get the initial external magnetic field strength, $B_{\mathrm{ex}}$, at this height. Since the quantity $c_0$ depends logarithmically on the ratio, $ {R_0} / {r_0} $, variation of the latter will not affect the former significantly. Therefore, we assign the same number of $ {R_0} / {r_0} $ as used by \cite{2006Kliem} for each simulation. Because it is hard to get the exact mass density of the flux rope, we assume the density of the order of magnitude of $10^{16}$ $\rm m^{-3}$, slightly varied with each inclination. As shown by Equation (\ref{equation}), all the aforementioned parameters do not significantly impact the acceleration profile computed from the model. Since we cannot determine the ejection direction from observations, we consider five cases of different inclination angles. For each direction, we carefully select the parameters, as listed in Table \ref{par}, to make the theoretical acceleration equivalent to the observed acceleration. The theoretical acceleration profiles are also transformed to that in the plane of sky. It is found that the acceleration curves of all the simulated cases are similar to the acceleration curve given by the third fitting, the best to the observed data (Figure~\ref{fig:cme_curve}(b) and~\ref{fig:cme_curve}(d)). Since the quantitative value of the simulated acceleration changes with different parameters, we further compute the derivative of the acceleration with respect to the propagation distance, ${d a}/{d R_{\rm dis}}$, where $a$ is the acceleration and $R_{\rm dis}$ is the distance in the plane of sky. The derivative curves are displayed by dashed lines in Figure~\ref{fig:cme_curve}. There appears a peak in the derivative curve for each inclination in the simulations, quite similar to what revealed by the third fitting to the observed data. More precisely, the derivative curve for an inclination of $45^\circ$ is the closest to that observed in both the shape of the curve and the height of the peak, which suggests that the flux rope may be ejected at this inclination. Even though the derivative curve in the $30^\circ$ case peaks around the same height, it decays too fast and becomes negative at 390$''$ (273 Mm) while the derivative value in the observation does not.


\section{Discussion and Summary } \label{sec:summary}
In this paper, we investigate the formation of the multiple ribbons of the X2.1 flare and the associated CME eruption on 2015 March 11 using the NLFFF model and the torus instability model. In particular, we study how the flare ribbons of different shapes are formed in terms of the specific magnetic topology and the related physical parameter, the squashing degree $Q$. We also compare the observed acceleration profile of the CME with that computed theoretically from the torus instability model to figure out the most likely ejection direction. The key findings are listed as follows:
\begin{enumerate}[1.]
\item The active region deviates from current neutralization, since the ratio $\rm |DC/RC|$ fluctuates in a range of 1.3--1.4 (1.5--1.7) in the positive (negative) polarities. This is consistent with the inference of \cite{2017Liu} that the CME-producing active regions are non-neutralized. The direct current is mainly distributed in the vicinity of the PILs and shows a coherent double-J-shaped pattern near the PIL between the smaller dipole N2 and P2. The return current is scattered around the periphery of the direct current, which is particularly obvious in the major positive polarity P1.
\item The extrapolated potential field reveals that the background magnetic field over a central flux rope weakens gradually with time, which may be the reason of the CME  eruption.
\item There is a large scale HFT as the intersection of a planar QSL and some arched ones crossing it. Underneath the HFT are two flux ropes. The central flux rope rises due to the weakening of the background magnetic field, resulting in two brightest parallel ribbons. During its rising, some slightly sheared lines around the flux rope reconnect sequentially forming an irregular ribbon and a J-shaped brightening. The erupting flux rope finally squeezes the large scale HFT causing magnetic reconnection there, producing a remote ribbon and a circular brightening. These flare ribbons are spatially consistent with some parts of the footprints of the QSLs on the boundary.
\item The flux rope seems to have two different phases of acceleration. It is firstly accelerated slowly and then undergoes a rapid acceleration when reaching a particular height. We also compute the theoretical acceleration profile based on the TI model with the observed decay index. If assuming an inclination of $45^\circ$, the theoretical acceleration curve is quite similar to the observed one.
\end{enumerate}

This flare has previously been studied by \cite{2016Li}, in which they focused on the quasi-periodic slipping reconnection tens of minutes prior to the eruption of the flare. They extrapolated the magnetic field of a smaller region that only exhibits the central flux rope, whose morphology is similar to that shown in our extrapolation. \cite{2016Li} thought that the flux rope loses its equilibrium because of the long-duration slipping reconnection occurring in the QSLs enveloping it. In another sense, the continuous reconnection of the neighbouring field lines might impair the background magnetic field above the flux rope. \cite{2015Joshi} reported a three-ribbon flare that comprises two parallel ribbons and a circular one, similar to the event studied here. They proposed that the flare is produced by the tether-cutting reconnection and subsequent null-point reconnection. The circular ribbon in that event is almost closed, which corresponds to the footprints of a large-scale fan-spine-type magnetic structure. However, in the event here, the southeastern part of the flare ribbon is still open since the center of the large scale HFT rises towards the southeast resulting in the separation of the footprints of the HFT in the southeast.

\cite{2006Kliem} presented a simplified TI model and performed simulations with various decay indices. In each simulation, however, the decay index stays constant that could be satisfied only in the case of small displacement. In reality, the decay index changes obviously with height \citep{1978vanTend}. In some cases, the height profile of decay index is saddle-shaped, as in the active region with multiple polarities \citep{2010Guo}. Thus, it is necessary to compute the acceleration of a CME with the observed decay index that is height-dependent. The result reveals that the most sensitive parameter to the theoretical acceleration profile is the ejection angle, which is undetermined from one-perspective observations. By adopting various ejection angles, we find an ejection angle of $45^{\circ}$ can best reproduce the acceleration profile as observed. This seems to be testified by the multi-wavelength observations of AIA, which shows that the flux rope propagates obliquely upwards. In fact, in the simulations of \cite{2010Aulanier}, the flux rope erupts towards the weaker polarity in an asymmetrical magnetic field. Likewise, in our case, the central flux rope is ejected at an inclination of  $45^\circ$ towards the northeast where the magnetic field is relatively weak.

In summary, we can understand the whole eruption process of the bifurcated flux rope FR1, combining the sequence of flare ribbons, the magnetic topology structures, and the acceleration profile of the CME. First, FR1 starts to rise up because of the weakening of the background field with time, causing magnetic reconnection beneath the flux rope. This results in the Lorentz force imbalance and in turn drives FR1 to rise higher. Then, the rising of FR1 squeezes the large scale HFT, making field lines in the vicinity of the HFT reconnect sequentially. As revealed by the acceleration profile, the rising driven by the Lorentz force imbalance induced by magnetic reconnection is close to a uniform acceleration process. When FR1 reaches the height where the decay index $n$ exceeds 1.5, the torus instability dominates the kinematic evolution of FR1 and drives it to ascend in an explosive way. Furthermore, the comparison between the theoretical acceleration and the actually observed one implies that FR1 is ejected in an inclination of $45^\circ$. There are still several issues that require further studies. First, is there a unified model for multiple-ribbon flares? Second, how to establish one-to-one correspondence between QSLs and flare ribbons? Or do we need to introduce other physical parameters (such as electric current) to constrain the relationship between QSLs and flare ribbons? Third, to what accuracy the simplified TI model can predict the acceleration of a CME occurring in a complex magnetic field like that in multi-ribbon flares.






\acknowledgments
 The authors are grateful to the anonymous referee for valuable comments. The data for this study were obtained by courtesy of NASA/\textit{SDO} and the AIA and HMI science teams. \textit{SDO} is the first launched mission of NASA’s Living With a Star Program. Y.Q., Y.G., M.D.D., and Z.Z. are supported by NSFC grants 11773016, 11733003, 11533005, and 11961131002. We are grateful to the High Performance Computing Center (HPCC) of Nanjing University for doing the numerical calculations in this paper on its blade cluster system.



\begin{table}[h]
\begin{threeparttable}
\begin{center}
\caption{The parameters for simulations of the CME acceleration.} \label{par}
\begin{tabular}{c c c c c c } \\
\hline \hline
Direction   &     $R_0~^{\rm a}$ (Mm)   &     $ {R_0} / {r_0} $    &     $c_{0}$    &   Number Density ($m^{-3}$)   &    $B_\mathrm{ex}$ (G) \\
\hline
$  30^\circ$   &   167.1  &  10   &  2.63   &   $4.0~\times~10^{16}$   &   5.3 \\
  $45^\circ$   &   158.5 &   10   &   2.63   &   $2.1~\times~10^{16}$   &  4.4  \\
  $60^\circ$   &   158.5 &   10   &   2.63   &   $1.0~\times~10^{16}$   &  3.8   \\
  $75^\circ$   &   165.7 &   10   &   2.63   &   $6.5~\times~10^{15}$   &  3.6   \\
  $90^\circ$   &   171.5 &   10   &   2.63   &   $5.5~\times~10^{15}$   &  3.9   \\
\hline
\end{tabular}
      \begin{tablenotes}
        \footnotesize
        \item[a] $R_0$ is the height at which the decay index reaches the threshold of 1.5 along each assumed ejection direction.
      \end{tablenotes}
\end{center}
\end{threeparttable}
\end{table}



\end{document}